\documentclass[10pt]{article}
\usepackage{amsmath,amsfonts}%,xypic}
\usepackage{amssymb}
\usepackage{amscd}
\usepackage{graphicx}
\usepackage{rotating}
\advance \textheight by \topskip
\makeatletter

\@addtoreset{equation}{section}
 
\makeatother
\newcommand{\beq}{\begin{equation}}
\newcommand{\eeq}{\end{equation}}
\newcommand{\beqa}{\begin{eqnarray}}
\newcommand{\eeqa}{\end{eqnarray}}
\newcommand{\noi}{\noindent}

\newcommand{\g}{{\mathfrak g}}

\def\>{\rangle}
\def\<{\langle}

\begin{document}

\title{{\bf  Kinematical superalgebras and Lie algebras of order $3$}  }
\author{{\sf  R. Campoamor-Stursberg}\thanks{e-mail:
rutwig@pdi.ucm.es}$\,\,$${}^{b}$ and {\sf  M. Rausch de
Traubenberg}\thanks{e-mail:
rausch@lpt1.u-strasbg.fr}$\,\,$${}^{b}$
\\
{\small ${}^{a}${\it IMI and Dpto. Geometr\'\i a y Topologia,
Fac. CC. Matem\'aticas,}} \\
{\small {\it Universidad Complutense de Madrid,
Plaza de Ciencias, 3, F-28040 Madrid, Spain}}  \\
{\small ${}^{b}${\it Laboratoire de Physique Th\'eorique, CNRS UMR
7085,Universit\'e Louis Pasteur}}\\
{\small {\it  3 rue de l'Universit\'e, 67084 Strasbourg Cedex,
France}} }
\date{ }
\maketitle

\vskip-1.5cm
%\vspace{2truecm}
\begin{abstract}
\end{abstract}
We study and classify kinematical algebras which appear in the
framework of Lie superalgebras or Lie algebras of order three. All
these algebras are related through generalised Inon\"u-Wigner
contractions from either the orthosymplectic superalgebra or the
de Sitter Lie algebra of order three.

\section{Introduction}
The theory of contractions of Lie groups entered physics providing
a formal derivation of classical mechanics from relativistic
mechanics by means of a limiting process between the underlying
symmetry groups. This approach allowed a precise interpretation of
the changes of physical generators during contraction, and allowed
also to obtain the Poincar\'e group as the limit of the de Sitter
groups, which constitute the only symmetry groups of General
Relativity having the same dimension. Bacry and L\'evy-Leblond
\cite{bl} systematically applied the contractions of Lie algebras
to classify the possible kinematical groups, basing on isotropy of
space and the following assumptions on the structure of
4-dimensional (homogeneous) space-time: (i) Time-reversal and
parity are automorphisms of the kinematical group, (ii)
Non-compactness of one-parametric subgroups generated by boost
transformations.

A remarkable fact on this classification is that all these
kinematical models arise as contractions of the real forms of the
rank two simple orthogonal Lie algebra $\frak{so}(5)$. In addition
to the well known Galilean algebra, another non-relativistic
kinematical group was found and studied in detail \cite{LL}. This
proved the versatility of contractions, allowing to explain
different models in a physically consistent manner, expressing
them through limits of the fundamental constants.\footnote{A
general scheme on possible physical limit procedures can be found
in \cite{Ke}.} The same idea of relating kinematical structures by
means of contractions has been successfully applied to more
general cases and situations, like superalgebras in the
supersymmetric frame, or more recently quantum groups
\cite{MdM,Hus}. In these constructions, the corresponding
Poincar\'e and Galilean algebras have been derived using the
supersymmetric and quantum versions of the de Sitter algebras,
extending in natural manner the classical kinematical frame. Most
of these works focus principally on the corresponding Poincar\'e
and Galilei models, leaving out the analysis of generalisations of
the non-standard kinematical groups like those of Carroll or
Newton type, which are more or less regarded as exotic models.

A lot of attention has been also devoted to non-relativistic
limits of supersymmetric theories \cite{Zu}, in analogy with the
classical Galilean relativity studied in \cite{LL1}. In \cite{Zu},
a systematic study of the Galilean limit of the superfield
formulation of the massive Wess-Zumino model and of supersymmetric
quantum electrodynamics (SQED) was developed. Criteria allowing
the definition of a super-Galilei group were obtained using the
notion of pseudo extension groups, emphasising the role of the mass
when expanding Galilean chiral superfields. It was already pointed
out there that in the massive case in higher dimensions, more than
one possible limit arises from the corresponding super-Poincar\'e
algebra, with a priori no distinguished contracted group.

\medskip
The main objective of this work is to extend the classical
kinematical classification of Bacry and L\'evy-Leblond to the
supersymmetric case and Lie algebras of order three, using the
same contraction ansatz, and including the  non-standard models of
Carroll and Newton. It will turn out that, with some exceptions,
for the superalgebras and Lie algebras of order three, similar
contraction diagrams to those found in \cite{bl} can be
obtained. \\

Lie algebras of order three (or more generally Lie algebras of
order $F$) were introduced in \cite{flie} as a possible
generalisation of Lie (super)algebras, in order to implement non-trivial
extensions of the Poincar\'e symmetry which are different than the
usual supersymmetric extension \cite{csusy}. A Lie algebra of
order $F$  admits a $\mathbb Z_F-$grading ($F=3$ in this paper),
the zero-graded part being a Lie algebra. An $F-$fold symmetric
product (playing the role of the anticommutator in the case $F=2$)
expresses the zero graded  part in terms of the non-zero graded
part. This new structure was subsequently applied within the
framework of the Poincar\'e algebra, and a Quantum Field Theory with
a  non-trivial extension, different from supersymmetry, was
explicitly constructed \cite{csusy}. Furthermore, the basis of the
theory of contractions and deformations in the context of Lie
algebras of order three has been studied in \cite{gr}.

\medskip

The notion of contraction of Lie algebras, although generally
given in terms of a representative of the isomorphism class,
follows more naturally from the geometry of orbits. Given a Lie
algebra $\frak{g}$ with structure tensor ${C_{ij}^{k}}$ over a
fixed basis $\left\{X_{i}\right\},i=1,..,n$, a linear redefinition
of the generators via a matrix $A\in GL(n,\mathbb{R})$ gives the
transformed structure tensor
\begin{equation}
{C^{\prime}}_{ij}^{
n}=A_{i}{}^{k}A_{j}{}^{\ell}(A^{-1})_{m}{}^{n}C_{k\ell}^{m}.
\label{BW}
\end{equation}

Taking into account all possible changes of basis, we obtain the
orbit $\mathcal{O}(\frak{g})$ of $\frak{g}$ by the action of the
general linear group $GL(n,\mathbb{R})$, consisting of all Lie
algebras that are isomorphic to $\frak{g}$. Thus, for describing
the Lie algebra, any of the elements (representatives) of the
orbit can be chosen. In this coordinate free interpretation, a Lie
algebra $\frak{g}^{\prime}$ is called a contraction of $\frak{g}$
if $\frak{g}^{\prime}$ belongs to the closure
$\overline{\mathcal{O}(\frak{g})}$ of the orbit. This naturally
leads to the traditional presentation in terms of limits.
Considering a family $\Phi_{\epsilon}\in \text{Aut}(\frak{g})$ of
non-singular linear maps of $\frak{g}$, where $\epsilon\in (0,1]$,
for any $X,Y\in\frak{g}$ we define
\begin{equation}
\left[X,Y\right]_{\Phi_{\epsilon}}:=\Phi_{\epsilon}^{-1}\left[\Phi_{\epsilon}(X),\Phi_{\epsilon}(Y)\right],
\end{equation}
which obviously reproduces the brackets of the Lie algebra over
the transformed basis. Actually this is nothing but equation
(\ref{BW}) for a special kind of transformations. Now suppose that
the limit
\begin{equation}
\left[X,Y\right]_{\infty}:=\lim_{\epsilon\rightarrow
0}\Phi_{\epsilon}^{-1}\left[\Phi_{\epsilon}(X),\Phi_{\epsilon}(Y)\right]
\label{Ko}
\end{equation}
exists for any $X,Y\in\frak{g}$. Then equation (\ref{Ko}) defines
a Lie algebra $\frak{g}^{\prime}$ which is a contraction of
$\frak{g}$, since it corresponds to a limiting point of the orbit.
We say that the contraction is non-trivial if $\frak{g}$ and
$\frak{g}^{\prime}$ are non-isomorphic, {\it i.e.}, if
$\frak{g}^{\prime}$ is a point of the frontier of
$\mathcal{O}(\frak{g})$, and trivial otherwise. In this sense
contractions should be understood as points of paths connecting
two representatives of the orbits of the corresponding Lie
algebras. As a consequence, two contractions
$\frak{g}_{1}\overset{f_{\epsilon}}{\longrightarrow}\frak{g}_{2}$
and
$\frak{g}_{1}^{\prime}\overset{g_{\epsilon}}{\longrightarrow}\frak{g}_{2}^{\prime}$
are equivalent if $\frak{g}_{1}^{\prime}\in
\mathcal{O}(\frak{g}_{1})$ and $\frak{g}_{2}^{\prime}\in
\mathcal{O}(\frak{g}_{2})$.

\medskip

A contraction for which there exists some basis
$\left\{Y_{1},..,Y_{n}\right\}$\footnote{The choice of basis
determines the structure constants, and therefore an orbit
representative of $\mathcal{O}(\frak{g})$} such that the
contraction matrix $A_{\Phi}$ is diagonal, that is, adopts the
form
\begin{equation*}
(A_{\Phi})_{ij}= \delta_{ij}\epsilon^{n_{j}},\quad
n_{j}\in\mathbb{Z},
\end{equation*}
is  called a generalised In\"on\"u-Wigner contraction. Although
there exist contractions that are not equivalent to this type,
which means that an arbitrary contraction does not necessarily
reduce to this form, for many physical problems they are of great
relevance. The diagonal transformations correspond to some scale
changes in the generators, and therefore have a precise physical
meaning. In was precisely in this sense that contractions were
introduced by Segal, In\"on\"u and Wigner \cite{IW}, in order to
describe a continuous transition from relativistic to
non-relativistic physics. These definitions can be generalised without effort
to superalgebras and other more general algebraic structures \cite{Ly}.

\medskip

In this work we focus primarily on generalised In\"on\"u-Wigner
(IW) contractions, for physical reasons. Since in the kinematical
frame the generators of the Lie algebra are identified with
physical operators, contractions obtained by re-scaling certain of
its generators still preserve this physical meaning, up to some
phase transitions for the rescaled elements. On the other hand,
the class of IW-contractions is the natural type of limiting
transformations related to semisimple Lie algebras and the
embedding of subalgebras \cite{C72}. Even if successive
composition of IW-contractions is not necessarily equivalent to a
general IW-contraction, in each step we deal with diagonal
transformations, which enables us to interpret how the symmetry
changes when modifying the main parameters.

The content of this paper is the following: In section two we
review the supersymmetric and Lie algebras of order three
extensions of the de Sitter algebra. Section three (respectively
four, five and six) are devoted to the study of extensions of the
Poincar\'e algebra (respectively to the extensions of  the Galilei
algebra,  of the Carroll algebra and of the Newton algebra).
Although there is only one physically consistent possible supersymmetric
extension of the corresponding Lie algebras, there are many
possible extensions of order three. Some of them
reproduce already known extensions considered in the
literature. It turns out that, in analogy to the Lie algebra case,
the kinematical superalgebras (resp. kinematical Lie algebras of
order three) are related though an IW-contraction of the
orthosymplectic superalgebra (resp. de Sitter algebra of order
three). Finally, more extended material related to the algebraic
definition of a Lie algebra of order three is given in the
appendix.

\section{Extensions of the (anti) de Sitter algebras}

We consider in  this section some possible extensions of the de
Sitter Lie algebras  $\mathfrak{so}(2,3)$ and
$\mathfrak{so}(1,4)$. Since the difference between these two
models is the signature of the metric tensor, the expressions
involving the generators can be developed without particular
reference to the metric. In the following, we just give the
algebraic structure for the anti-de Sitter case
$\mathfrak{so}(2,3)$, the argument being similar for the remaining
model by simply replacing the metric tensor. We consider the usual
basis $\left<L_{MN}=-L_{NM}, 0\le M < N \le 4\right>$ with
commutation relations \beqa \label{so23} \left[L_{MN},
L_{PQ}\right]&=& \eta_{NP} L_{MQ} - \eta_{MP} L_{NQ} + \eta_{QN}
L_{PM} - \eta_{QM} L_{PN}, \eeqa

\noi where $\eta_{MN}=\text{diag}(1,-1,-1,-1,1)$. It follows at
once that the generators $L_{MN}$ with $M,N\neq 4$ span the
Lorentz algebra $\frak{so}(1,3)$. In the basis $<L_{\mu
\nu},P_\mu=L_{\mu 4}, \ \mu,\nu=0,\cdots,3>$ the commutation
relations are rewritten as
\beqa \label{so23L} \left[L_{\mu \nu}, L_{\rho \sigma}\right]&=&
\eta_{\nu \rho} L_{\mu \sigma} - \eta_{\mu \rho} L_{\nu \sigma} +
\eta_{\sigma \nu} L_{\rho \mu} - \eta_{\sigma \mu}
L_{\rho \nu},\nonumber\\
\left[L_{\mu \nu}, P_{\rho}\right]&=& \eta_{\nu \rho } P_\mu -
\eta_{\mu \rho}
P_\nu,\nonumber\\
\left[P_\mu, P_\nu\right]&=& L_{\nu \mu}. \eeqa
For later use, when dealing with contractions of $\frak{so}(2,3)$,
we also rewrite this algebra in the Bacry-L\'evy-Leblond notation:
We rename the basis elements as $K_{i}=L_{0i}$, $P_{i}=L_{i4}$,
$H=L_{04}$ and $L_{i}=L_{jk}$, where $i,j,k$ are taken in cyclic
order. From now on, any relation like $[L_i,L_j]= L_k$ means that
$i,j,k$ are taken in cyclic order.
 The brackets of the anti-de Sitter algebra over this basis are:
\beqa \label{so23bis}
\begin{array}{llll}
\left[L_{i}, L_{j}\right]= L_{k},& \left[L_{i}, K_{j}\right]=
K_{k},& \left[L_{i}, P_{j}\right]=
P_{k},& \left[K_{i}, K_{j}\right]= -L_{k},\nonumber \\
\left[P_{i}, P_{j}\right]= -L_{k},& \left[K_{i}, P_{j}\right]=
-\delta_{i j}H,& \left[K_{i}, H\right]=-P_{i},& \left[P_{i},
H\right]= K_{i}.
\end{array}
\eeqa

In order to shorten notations, we will write
$\left[\mathbf{L,L}\right]=\mathbf{L}$ for
$\left[L_{i},L_{j}\right]=L_{k}$. As proved in \cite{bl}, all
classical kinematical algebras arise as In\"on\"u-Wigner
contractions of the de Sitter algebras. The brackets of the
contractions are reproduced in Table 1.

\begin{table}[h]
\caption{Non-vanishing brackets of classical kinematical algebras
in the standard basis. The common brackets to all Lie algebras
below are those corresponding to space isotropy:
$\left[\mathbf{L,L}\right]=\mathbf{L}$,
$\left[\mathbf{L,K}\right]=\mathbf{K}$ and
$\left[\mathbf{L,P}\right]=\mathbf{P}$.}
\begin{tabular}
[c]{l|cccccccccc}\hline\hline & $\frak{so}\left(  2,3\right)  $ &
$\frak{so}\left(  1,4\right) $ & $I\frak{so}\left(  1,3\right)  $
& $I\frak{so}\left( 4\right) $ & $I\frak{so}\left(  1,3\right)
^{\prime}$ & Carroll & $Ne^{\exp}$ & $Ne^{osc}$ & $G\left(
2\right)  $ & $G\left( 2\right) ^{\prime}$\\\hline
 $\left[  \mathbf{K,K}\right]  $ & $-\mathbf{L}$ & $-\mathbf{L}$ &
$-\mathbf{L}$ & $0$ & $0$ & $0$ & $0$ & $0$ & $0$ & $0$\\
$\left[  \mathbf{K,P}\right]  $ & $-H$ & $-H$ & $-H$ & $-H$ & $-H$
& $-H$ &$0$ & $0$ & $0$ & $0$\\
$\left[  \mathbf{P,P}\right]  $ & $-\mathbf{L}$ & $\mathbf{L}$ &
$0$ &$\mathbf{L}$ & $-\mathbf{L}$ & $0$ & $0$ & $0$ & $0$ & $0$\\
$\left[  \mathbf{K,}H\right]  $ & $-\mathbf{P}$ & $-\mathbf{P}$ &
$-\mathbf{P}$ & $0$ & $0$ & $0$ & $-\mathbf{P}$ & $-\mathbf{P}$ &
$-\mathbf{P}$ & $0$\\
$\left[  \mathbf{P,}H\right]  $ & $\mathbf{K}$ & $-\mathbf{K}$ &
$0$ & $-\mathbf{K}$ & $\mathbf{K}$ & $0$ & $-\mathbf{K}$ &
$\mathbf{K}$ & $0$ & $-\mathbf{K}$\\\hline\hline
\end{tabular}
\end{table}

It is straightforward to verify that the Poincar\'e and
Para-Poincar\'e algebras ($I\frak{so}(1,3)$ and
$I\frak{so}(1,3)^{\prime}$, respectively) are isomorphic as Lie
algebras, though they are physically different. The same happens
with the two Galilei algebras. The physical distinction follows
from the interpretation of the generators. For this reason, from
now on, we will only consider the Poincar\'e and Galilei algebras,
the computations being completely analogous for the Para-models by
simply interchanging the generators.

\subsection{The $\mathfrak{osp}(1|4)$ algebra}
To construct a supersymmetric extension of the (anti-)de Sitter
algebra, we start from the real forms of the orthosymplectic
superalgebra $\mathfrak{osp}(1|4,\mathbb C) =
\mathfrak{sp}(4,\mathbb C) \oplus \mathbb C^4$. Among the Lie
algebras $\mathfrak{so}(2,3)$ and $\mathfrak{so}(1,4)$, only the
former admits a four-dimensional real spinor  representation (the
Majorana spinors ), therefore only $\mathfrak{so}(2,3)$ will admit a
supersymmetric extension. Using the natural inclusion
$\mathfrak{so}(1,3) \subset \mathfrak{so}(2,3)$ in the basis
(\ref{so23}), we consider the following decomposition
$\mathfrak{osp}(1|4)= \mathfrak{so}(2,3) \oplus \mathbb R^4 =
\left\<L_{\mu \nu}, P_\mu\right> \oplus \left< S_\alpha, \bar
S^{\dot \alpha}, \alpha, \dot \alpha =1,2 \right>$, where
$(S_\alpha, \bar S^{\dot \alpha})$ is a four dimensional Majorana
spinor ($\bar S^{\dot\alpha}{}^\star = S^\alpha$, where $^\star$
denotes the complex conjugation) of $\mathfrak{so}(1,3)$. Now,
taking for the Dirac $\Gamma-$ matrices
$$\Gamma_M=\left\{
\begin{array}{lll}
\Gamma_\mu&=&\begin{pmatrix} 0&\sigma_\mu \\
                           \bar \sigma_\mu& 0 \end{pmatrix} \\
\Gamma_4&=&\begin{pmatrix} 1&0 \\
                           0& -1 \end{pmatrix}
\end{array} \right.
$$

\noi with $\sigma_\mu=(1,\sigma_i), \bar \sigma_\mu =
(1,-\sigma_i)$, the $\sigma_i$ ($i=1,2,3$) denoting the Pauli spin
matrices. The index structure of the $\sigma_\mu-$matrices is as
follows: $\sigma_\mu \to \sigma_\mu{}_{\alpha \dot \alpha},\; \bar
\sigma_\mu \to \bar \sigma_\mu{}^{\dot \alpha  \alpha}$. We also
define $S_\alpha =\varepsilon_{\alpha\beta}S^\beta$, $S^\alpha
=\varepsilon^{\alpha\beta}S_\beta$, $\bar S_{\dot\alpha}=\bar
\varepsilon_{\dot\alpha \dot\beta}\bar S^{\dot\beta}$, $\bar
S^{\dot\alpha} =\bar \varepsilon^{\dot\alpha\dot\beta}\bar
S_{\dot\beta}$   with $\varepsilon, \bar \varepsilon$
antisymmetric matrices given by $\varepsilon_{12} = \bar
\varepsilon_{\dot 1\dot 2}=-1$ and $\varepsilon^{12} = \bar
\varepsilon^{\dot 1\dot 2}=1$, respectively. This means that if we define
$$
S_A= \begin{pmatrix} S_\alpha \\ \bar S^{\dot \alpha}
\end{pmatrix},
$$

\noi a Majorana spinor of $\mathfrak{so}(1,3)$, we have that
$$S^A= C_4{}^{AB} S_B, \ \ \ C_4^{AB}=
\begin{pmatrix} \epsilon^{\alpha \beta}&0 \\
       0&\bar  \epsilon_{\dot \alpha \dot \beta}
\end{pmatrix},$$
and $C_4$ defines a metric on the spinor space, seen as a $(1+3)D
$-spinor (see {\it e.g.} \cite{clif}). Introducing now the
$\mathfrak{so}(2,3)$ generators in the spinor representation
$$\Gamma_{MN}=\frac14\left(\Gamma_M \Gamma_N - \Gamma_N \Gamma_M\right)=
\left\{ \begin{array}{lllll}
\Gamma_{\mu \nu} &=&\begin{pmatrix} \sigma_{\mu \nu}&0 \\
                                    0& \bar \sigma_{\mu \nu}
\end{pmatrix}
&=&
\begin{pmatrix} \frac14(\sigma_\mu \bar \sigma_\nu -\sigma_\nu \bar \sigma_\mu)&0\\
                       0&  \frac14(\bar \sigma_\mu  \sigma_\nu -\bar \sigma_\nu  \sigma_\mu)\\
\end{pmatrix} \\
\Gamma_{\mu 4} &=& \begin{pmatrix} 0& -\frac12 \sigma_\mu \\
                                  \frac12 \bar \sigma_\mu & 0
\end{pmatrix}
\end{array} \right.
$$

\noi the index structure of the $\sigma_{\mu \nu}-$matrices is
reformulated as $\sigma_{\mu \nu} \to \left(\sigma_{\mu \nu}
\right)_\alpha{}^\beta,\; \bar \sigma_{\mu \nu} \to \left(\bar
\sigma_{\mu \nu} \right)^{\dot \alpha}{}_{\dot \beta}$. To
construct the orthosymplectic superalgebra $\mathfrak{osp}(1|4)$,
we notice at first that if we consider now $S_A$ as an
$\mathfrak{so}(2,3)$ Majorana spinor, the spinor metric is given
by
$$
C_5^{AB}=
\begin{pmatrix} \epsilon^{\alpha \beta}&0 \\
       0&-\bar  \epsilon_{\dot \alpha \dot \beta}
\end{pmatrix},
$$
and, with respect to $C_5$ $\Gamma_{MN}{}_A{}^B C_5{}_{DB}$, are
symmetric matrices (see for instance \cite{clif}). Now, since our
conventions are such that the R.H.S. of all commutators for a Lie
algebra have no pure imaginary factor $i$ ({\it i.e.} the
structure constants are real), for an unitary representation the
generators of the algebra are given by antihermitian  operators
(this convention is different with respect to that usually considered in physics).
This means that in order to construct
the orthosymplectic algebra with real structure
constants, since in the Majorana representation
the $\Gamma_{MN}$ matrices are real, we need to consider
generators in the fermionic sector which respect  this
convention: we will use rescaled  $S_A$'s such that $S_\alpha^\star
=i \bar S_{\dot \alpha}$ and $\bar S_{\dot \alpha}^\star = i
S_\alpha$. With this notation, the $\mathfrak{osp}(1|4)$ algebra
reads\footnote{In the convention of \cite{wb} the generators of the bosonic
and fermionic part of the algebra are hermitian and thus the structure
constants for all the commutators (resp. anticommutators) are purely
imaginary (resp. real) although in the conventions of \cite{w} the generators
of the bosonic (resp. fermionic) part of the algebra are anti-hermitian
(resp. hermitian) hence the structure constant for
all the commutators (resp. anticommutators) are
real (resp. purely imaginary). With our conventions, the structure constant
of both the commutators and anticommutators are real.}
$$
\left\{S_A, S_B\right\} = b\ \Gamma_{MN}{}_A{}^D C_5{}_{BD}
L^{MN},
$$
\noi with $b$ positive or $b$ negative.
The two choices for the sign of $b$ correspond to the two
real forms of $\mathfrak{osp}(1|4, \mathbb C)$. In the
$\mathfrak{so}(1,3)$ notations, the algebra takes the form (we
have  chosen here the real form corresponding to $b>0$ and we have
normalised the generators such that $b=4$)
\beqa \label{osp} &&\left[L_{\mu \nu}, S_\alpha\right]=
(\sigma_{\mu \nu})_\alpha{}^\beta S_\beta, \ \ \left[L_{\mu4}
S_\alpha\right]= -\frac12 \sigma_{\mu \alpha \dot \alpha} \bar
S^{\dot \alpha},\ \ \left[L_{\mu \nu}, \bar S^{\dot
\alpha}\right]= (\bar \sigma_{\mu \nu})^{\dot \alpha}{}_{\dot
\beta} \bar S^{\dot \beta}, \ \ \left[L_{\mu4}, \bar  S^{\dot
\alpha}\right]= \frac12 \bar \sigma_{\mu}^{\dot  \alpha  \alpha}
S_{ \alpha},
\nonumber \\ \nonumber \\
&&\left\{S_\alpha, S_\beta\right\}= 4(\sigma^{\mu \nu})_{\alpha
\beta} L_{\mu \nu}, \ \left\{\bar S^{\dot \alpha}, \bar S^{\dot
\beta}\right\}= -4(\bar \sigma^{\mu \nu})^{\dot \alpha \dot \beta}
L_{\mu \nu},\ \left\{S_\alpha, \bar S_{\dot \beta}\right\}=
2 (\sigma^\mu)_\alpha{}_{\dot \beta} P_\mu. \nonumber \eeqa

\noi together with those of \eqref{so23L}.

\subsection{The de Sitter algebra of order 3}

With the previous notations, we are in situation of constructing
an anti-de Sitter algebra of order 3. (See the Appendix for the
definition of  Lie algebras of order three.) To this extent, we
consider the following Lie algebra decomposition: $\g =
\mathfrak{so}(2,3) \oplus \text{ad } \mathfrak{so}(2,3) =
\left<L_{MN}=-L_{NM}, 0\le M < N \le 4\right> \oplus
 \left<A_{MN}=-A_{NM}, 0\le M < N \le 4\right>$\footnote{We could
have considered, on an equal footing, the Lie algebra of order 3
$\g = \mathfrak{so}(1,4) \oplus \text{ad } \mathfrak{so}(1,4)$.}
with additional bilinear and trilinear brackets:
\beqa \left[L_{MN}, A_{PQ}\right]&=& \eta_{NP} A_{MQ} - \eta_{MP}
A_{NQ}
+ \eta_{QN} A_{PM} - \eta_{QM} A_{PN}, \\
\left\{A_{MN}, A_{PQ}, A_{RS}\right\}&=& (\eta_{MP} \eta_{NQ} -
\eta_{MQ} \eta_{NP}) L_{RS} + (\eta_{MR} \eta_{NS} - \eta_{MS}
\eta_{NR}) L_{PQ}+ (\eta_{PR} \eta_{QS} - \eta_{PS} \eta_{QR})
L_{MN} \nonumber \eeqa

\noi together with the relations \eqref{so23}. Using the
$\mathfrak{so}(1,3)$-basis previously mentioned, $\text{ad }
\mathfrak{so}(2,3)$ is generated by $A_{\mu \nu}, C_\mu = A_{\mu
4}$, $\mu, \nu =0,1,2,3$, and the algebra reads \beqa
\label{3-AdS}
 \left[L_{\mu \nu}, A_{\rho \sigma}\right]&=& \eta_{\nu \rho }
A_{\mu \sigma} - \eta_{\mu \rho } A_{\nu \sigma}
+ \eta_{\sigma \nu } A_{\rho \mu } - \eta_{\sigma \mu } A_{\rho \nu}, \nonumber \\
\left[L_{\mu \nu}, C_\rho\right]&=&\eta_{\nu \rho} C_\mu
-\eta_{\mu \rho} C_\nu,
\nonumber \\
\left[P_\mu, A_{\nu \rho}\right]&=&
-\eta_{\rho \mu} C_\nu + \eta_{\nu \mu} C_\rho, \nonumber \\
\left[P_\mu, C_\nu\right]&=& A_{\nu \mu}.  \nonumber \\
%%%%%%%%%%%%%%%
\\
\left\{A_{\mu \nu }, A_{\pi \kappa}, A_{\rho \sigma}\right\}&=&
(\eta_{\mu \pi } \eta_{\nu \kappa} - \eta_{\mu \kappa} \eta_{\nu
\pi}) L_{\rho \sigma} + (\eta_{\mu \rho } \eta_{\nu \sigma} -
\eta_{\mu \sigma} \eta_{\nu \rho}) L_{\pi \kappa}+
(\eta_{\pi \rho } \eta_{\kappa \sigma} - \eta_{\pi \sigma} \eta_{\kappa \rho}) L_{\mu \nu}, \nonumber \\
\left\{A_{\mu \nu }, A_{\pi \kappa}, C_{\rho }\right\}&=&
(\eta_{\mu \pi } \eta_{\nu \kappa} - \eta_{\mu \kappa} \eta_{\nu \pi}) P_{\rho},\nonumber \\
\left\{A_{\mu \nu }, C_{\rho }, C_{\sigma }\right\}&=&
\eta_{\rho \sigma} L_{\mu \nu}, \nonumber \\
\left\{C_\mu, C_\nu, C_\rho\right\}&=& \eta_{\mu \nu} P_\mu +
\eta_{\mu \rho} P_\nu + \eta_{\nu \rho} P_\mu, \nonumber \eeqa

\noi together with the relations \eqref{so23L}. This algebra was
first introduced in \cite{flie}, where it has been established
that to any Lie (super)algebra we can associate a Lie algebra of
order $F$ structure by means of an induction based argument. The
algebra \eqref{3-AdS} appears as a special case of this induction
theorem.

\section{Extensions of the Poincar\'e algebra}
In the classical frame, the Poincar\'e algebra is obtained from
the anti-de Sitter algebra through the Inon\"u-Wigner contraction
defined by the transformations
\beqa \label{IW-Poin} L'_{\mu \nu}=L_{\mu \nu}, \ \ P'_\mu
=\varepsilon P_\mu \eeqa
\noi and taking the limit $\varepsilon \to 0$. The non-vanishing
brackets are \beqa \label{poin}
 \left[L'_{\mu \nu}, L'_{\rho \sigma}\right]&=& \eta_{\nu \rho }
L'_{\mu \sigma} - \eta_{\mu \rho } L'_{\nu \sigma}
+ \eta_{\sigma \nu } L'_{\rho \mu } - \eta_{\sigma \mu } L'_{\rho \nu}, \nonumber \\
\left[L'_{\mu \nu}, P'_\rho\right]&=&\eta_{\nu \rho} P'_\mu
-\eta_{\mu \rho} P'_\nu,
\nonumber \\
 \left[P'_\mu, P'_\nu\right]&=&0. \eeqa

It is natural to ask whether a similar procedure holds for the
supersymmetric extensions. Due to isomorphism of the anti-de
Sitter algebra with the non-compact symplectic algebra
$\frak{sp}(4,\mathbb{R})$, the correct choice for the contraction
in the supersymmetric case is the orthosymplectic algebra
$\mathfrak{osp}(1|4)$ instead of the superalgebras
$\mathfrak{osp}(5|N)$\footnote{These algebras violate the
physically supersymmetric principle of requiring that the elements
of the Fermi sector transform as Lorentz spinors.} \cite{freu}.

In addition to the generators of \eqref{IW-Poin}, we consider also
transformed generators in the Fermi sector of
$\mathfrak{osp}(1|4)$ defined by \beqa Q_\alpha = \varepsilon^a
S_\alpha, \ \ \bar Q^{\dot \alpha} = \varepsilon^a \bar S^{\dot
\alpha}. \eeqa

\noi This contraction pattern is consistent, {\it i.e.}, the limit
exists for $\varepsilon\rightarrow 0$ and defines a superalgebra,
if the condition $2 a \ge 1$ is satisfied. For the value
$a=\frac12$ we find the algebra
\beqa \label{susy} \left[L'_{\mu \nu} Q_\alpha\right]&=&
(\sigma_{\mu \nu})_\alpha{}^\beta Q_\beta, \ \
\left[P'_\mu,Q_\alpha\right]=0,
\nonumber \\
\left[L'_{\mu \nu} \bar Q^{\dot \alpha}\right]&=& (\bar
\sigma_{\mu \nu})^{\dot \alpha}{}_{\dot \beta} \bar  Q^{\dot
\beta}, \ \ \left[P'_\mu,\bar Q^{\dot \alpha}\right]=0,
 \\
\left\{Q_\alpha, Q_\beta\right\}&=&0, \ \ \left\{\bar Q^{\dot
\alpha},\bar Q^{\dot \beta} \right\}=0, \ \ \left\{Q_\alpha, \bar
Q_{\dot \beta}\right\}= 2 \sigma^\mu{}_{\alpha \dot \beta }
P'_\mu, \nonumber \eeqa

\noi which is the well-known supersymmetric algebra. In the
kinematical notations of Table 1, these brackets reduce to
\beqa &&\left[L'_k, Q_\alpha\right]= -\frac{i}{2}
(\sigma_k)_\alpha{}^\beta Q_\beta, \ \left[K'_k, Q_\alpha\right]=
-\frac{1}{2} (\sigma_k)_\alpha{}^\beta Q_\beta, \nonumber \\
&&\left[L'_k, \bar Q^{\dot \alpha}\right]= -\frac{i}{2}
(\sigma_k)^{\dot \alpha}{}_{\dot \beta}
 \bar Q^{\dot \beta}, \ \left[K'_k, \bar Q^{\dot \alpha}\right]=
\frac{1}{2} (\sigma_k)^{\dot \alpha}{}_{\dot \beta}
\bar  Q^{\dot \beta}, \nonumber \\
&&\left\{Q_\alpha, \bar Q_{\dot \beta}\right\}= 2
\sigma_i{}_{\alpha \dot \alpha} P'_i + 2 \sigma_0{}_{\alpha
\dot \beta} H'. \eeqa

\noi Here we have $\sigma_i{}_{\alpha \dot \alpha} \bar
\sigma_j{}^{\dot \alpha \beta}=-i(\sigma_k)_\alpha{}^\beta$, $\bar
\sigma_i{}^{\dot \alpha  \alpha}  \sigma_j{}_{ \alpha \dot
\beta}=-i(\sigma_k)^{\dot \alpha}{}_{\dot \beta}$ (with $i,j,k$ in
circular permutation) or  $(\sigma_i)_\alpha{}^\beta=-
\sigma_0{}_{\alpha \dot \alpha} \bar \sigma_i{}^{\dot \alpha
\beta}$ and $(\sigma_i)^{\dot \alpha}{}_{\dot \beta}= \bar
\sigma_0{}^{\dot \alpha \alpha} \sigma_i{}_{\alpha \dot \beta}$.

This supersymmetric extension of the Poincar\'e algebra was first
proposed by Gol'fand and Likhtman in 1971, although its formal
introduction in physics is due to Wess and Zumino \cite{GL}. This
algebra moreover arises as a contraction of the orthosymplectic
superalgebra $\mathfrak{osp}(1|4)$, thus leads to a consistent
contraction pattern for supersymmetric extensions of kinematical
algebras.

There is however another possibility of contracting the
$\mathfrak{osp}(1|4)$ superalgebra such that it is still
consistent with the Poincar\'e algebra. This alternative was
introduced in \cite{Ko}, and proven to be an extension of the
Poincar\'e algebra by odd twistorial generators. Since for this
extension no hermitian representations for the generators in the
Fermi part exist, the so-called Konopel'chenko algebra is of no
use in supersymmetry.\footnote{Formally both Lagrangian are quite
similar, but the pure complex character for the Konopel'chenko
extension prevents it from providing physically realistic
descriptions.} Because of the non-reality of the latter algebra,
we keep it out from our construction of Lie algebras of order
three.

\subsection{Cubic Poincar\'e algebras}

Following the previous pattern of obtaining a supersymmetric
extension of the Poincar\'e algebra by means of generalisations of
the classical kinematical and supersymmetric contraction, we
search for cubic extensions of the Poincar\'e algebra that can be
obtained contracting the corresponding anti de Sitter algebra of
order three. To this extent, in addition to \eqref{IW-Poin} we
consider the following re-scaled generators:

\beqa \label{fpoin} U_{\mu \nu} = \varepsilon^a A_{\mu \nu}, \ \
V_\mu = \varepsilon^b C_\mu. \eeqa

\noi Expressing the binary and ternary brackets in this basis and
taking the limit $\varepsilon \to 0$, we find that this
contraction pattern is consistent only if the following system on
$a,b$ is satisfied\footnote{By consistent contraction pattern we mean that the corresponding
limit \eqref{Ko} exists for any pair and triple of generators.}:
\beqa \label{ine-fpoin} 1 + a - b \ge 0,\ 1 + b -a \ge 0,\ 3a\ge
0,\ 2a +b-1 \ge 0,\ a +2b\ge 0,\ 3b -1 \ge 0. \eeqa

We observe that, if all inequalities are strict, then all involved brackets
vanish and the corresponding extension is trivial. In order that a bilinear
or trilinear is preserved, it is necessary that the corresponding inequality
on the contraction parameters is an equality. Therefore, the non-trivial extensions
are given by those solutions of the preceding system for which at least one equality
is satisfied. The resulting extensions with non-vanishing trilinear bracket are given schematically
in Table 2.

\begin{table}[h]
\caption{Poincar\'e algebras of order three}
\begin{center}
\begin{tabular}{|l||l|c|c|c|c|c|c|}
\hline
&  & $[P^{\prime },U]$ & $[P^{\prime },V]$ & $\{U,U,U\}$ & $\{U,U,V\}$ & $%
\{U,V,V\}$ & $\{V,V,V\}$ \\ \hline 1. &
$a=0,b=1$%
& $V$ & $0$ & $L$ & $P$ & $0$ & $0$ \\ \hline 2. & $1+a-b=0$ & $V$
& $0$ & $0$ & $0$ & $0$ & $0$ \\
\hline 3. & $a=\frac43,\;b=\frac13$ & $0$ &
$U$ & $0$ & $0$ & $0$ & $P$ \\ \hline 4. &
$1+b-a=0$%
& $0$ & $U$ & $0$ & $0$ & $0$ & $0$ \\
\hline 5. & $2a+b-1=0$ &
$0$ & $0$ & $0$ & $P$ & $0$ & $0$ \\
\hline 6. &
$a=b=\frac{1}{3}$%
& $0$ & $0$ & $0$ & $P$ & $0$ & $P$ \\
\hline 7. & $b=\frac{1}{3}$
& $0$ & $0$ & $0$ & $0$ & $0$ & $P$ \\ \hline
\end{tabular}
\end{center}
\end{table}

\medskip

\noi The precise form of the brackets can be read off from
\eqref{fpoin}, {\it e.g.} $\{V,V,V\}=P'$ means that
$\{V_\mu,V_\nu,V_\rho\}=\eta_{\mu \nu} P'_\rho + \eta_{\mu \rho}
P'_\nu  + \eta_{\nu \rho} P'_\mu$, and the brackets $[L',U]$ and
$[L',V]$ are always different from zero. The extensions where
$[P',U]$ and/or $[P',V]$ are different from zero
are interesting since the translations are
represented by non-vanishing matrices. These are special cases of
Lemma 1 of \cite{gr}. Furthermore, there  is only one cubic
algebra which admits the subalgebra generated by
$\left<L'_{\mu,\nu}, P'_\mu\right> \oplus \left< V_\mu\right>$,
which corresponds to the cubic extension studied in \cite{csusy}.
This cubic extension of the Poincar\'e algebra has been
implemented in Quantum Field Theory \cite{csusy}. Expressed in terms
of the kinematical basis $\left<L'_i, K'_i, P'_i,H'\right> \oplus
\left<V_i, W=V_0\right>$, it adopts the form

\beqa
\label{3poin}
\begin{array}{ll}
\left[L'_i, V_j\right]=V_k,&
\left[K'_i, V_j\right]=-\delta_{ij} W, \\
\left[L'_i, W\right]=0,&
\left[K'_i, W\right]=-V_i\\
\left\{V_i,V_j,V_k\right\}= -\delta_{ij} P'_k -\delta_{ik} P'_j -
\delta_{jk} P'_i &
\left\{V_i,V_j,W\right\}= -\delta_{ij} H', \\
\left\{V_i,W,W\right\}= P'_i, & \left\{W,W,W\right\}=3 H'.
\end{array}
\eeqa

In the contraction scheme above, the generators before
contraction are denoted by unprimed symbols, while those after contraction
are primed. From now on,  for any contractions of the Poincar\'e algebra, the
generators of $I\mathfrak{so}(1,3)$ will be unprimed while
those of the contracted algebra will be primed. Since the Carroll
and the Galilei algebras are obtained through an Inon\"u-Wigner
contraction from the Poincar\'e algebra, we will only consider the
Galilei and Carroll algebras of order three related to the
Poincar\'e algebra of order three \eqref{3poin} {\it i.e.}
generated by $<L_i,K_i,P_i,H> \ \oplus \ <V_i, W>$. This choice is justified
by the interest of the algebra \eqref{3poin} in the quantum field theoretic approach,
while for the remaining cubic extensions (algebras 1-6 of Table 2) no consistent model
has been developed yet, thus not allowing to interpret in some physical manner the resulting
contractions.

\section{Extensions of the Galilei algebra}
The Galilei algebra is obtained from the Poincar\'e algebra
$I\frak{so}(1,3)$ through the contraction defined by the
transformations
\beqa \label{IW-Gal} L'_i=L_i, \ \ K'_i=\varepsilon K_i, \ \ P'_i=
\varepsilon P_i, \ \ H'=H. \eeqa

\noi The brackets are given in Table 1. In order to construct a
super-Galilei algebra by means of contractions, we add to the
previous generators the following ones:
\beqa \label{IW-supergal}
Q'_\alpha = \epsilon^a Q_\alpha, \ \ \bar Q'_\alpha = \epsilon^b
\bar Q_\alpha.
\eeqa

\noi In the inclusion $\mathfrak{so}(3) \subset
\mathfrak{so}(1,3)$, the two representations $\left<
Q_\alpha\right>$ and $\left< \bar Q_{\dot \alpha}\right>$ become
equivalent (but complex conjugate), we thus denote them as $\bar Q_{\dot
\alpha} \to \bar Q_{\alpha}$ and $\sigma_i{}_{\alpha \dot \beta}
\to \sigma_i{}_{\alpha \beta}$. The contraction scheme above leads
to a superalgebra whenever the condition $a + b -1 \ge 0$ is
satisfied. For the special case $a+b-1=0$ we get the $N=2$
supersymmetric extension (without central charge)
of the Galilei algebra given by:

\beqa
\label{sugal}
\begin{array}{cc}
\begin{array}{cc}
\left[L'_k, Q'_\alpha\right]= -\frac{i}{2}
(\sigma_k)_\alpha{}^\beta Q'_\beta,& \left[K'_k,
Q'_\alpha\right]=0, \\
\left[L'_k, \bar Q'_\alpha\right]= -\frac{i}{2}
(\sigma_k)_\alpha{}^\beta \bar Q'_\beta,& \left[K'_k,
\bar Q'_\alpha\right]=0,
\end{array}
& \left\{Q'_\alpha, \bar Q'_\beta\right\}= 2
\sigma_{i}{}_{\alpha \beta} P^{\prime i},
\end{array}
\eeqa

It should be remarked that, like happens with the Poincar\'e case,
this is not the only possibility for a supersymmetric Galilei
algebra. There is another possible contraction taking into account
a special $\mathbb{Z}_{4}$-grading of the Poincar\'e superalgebra
\cite{Hus}. The authors are however not aware of supersymmetric
field models based on this second Galilei superalgebra, which
justifies that it is not considered further to obtain the
corresponding Lie algebras of order three.

\subsection{Galilei algebras of order 3}

To construct the Galilei algebras of  order three from the cubic
extension of Poincar\'e \eqref{3poin}, we consider the following
generators \beqa \label{IW-3gal} V'_i=\epsilon^a V_i, \ \
W'=\epsilon^b W, \eeqa

\noi in addition to those of \eqref{IW-Gal}. The constraints to be
satisfied by the powers $a,b$ to define consistent contractions
are (see \eqref{3poin}): \beqa \label{ine-3gal} 1+a-b \ge 0,
1+b-a\ge 0,\ a \ge \frac{1}{3}, 2a+ b \ge 0,\ a + 2b -1 \ge 0,\
b\ge 0. \eeqa \noi The solutions to these equations provide the
possible non-equivalent cubic extensions of the Galilei algebra.
In any case, we have the common bracket
$\left\{V',V',W'\right\}=0$. The remaining brackets of the
different extensions are given schematically in Table 3.

\begin{table}[h]
\caption{Galilei algebras of order three}
\begin{center}
\begin{tabular}{|l||l|c|c|c|c|c|c|}
\hline &  & $\left[ K^{\prime },V^{\prime }\right] $ & $\left[
K^{\prime },W^{\prime }\right] $ & $\left\{ V^{\prime },V^{\prime
},V^{\prime
}\right\} $ & $\left\{ V^{\prime },V^{\prime },W^{\prime }\right\} $ & $%
\left\{ V^{\prime },W^{\prime },W^{\prime }\right\} $ & $\left\{
W^{\prime },W^{\prime },W^{\prime }\right\} $ \\ \hline 1. &
$a=\frac13, b=\frac43$
& $W^{\prime }$ & $0$ & $P^{\prime }$ & $0$ & $0$ & $0$ \\ \hline
2. & $1+a-b=0$ & $W^{\prime }$ & $0$ & $0$ & $0$ & $0$ & $0$ \\
\hline
3. & $a=1,b=0$ & $0$ & $V^{\prime }$ & $0$ & $0$ & $P^{\prime }$ & $%
H^{\prime }$ \\ \hline 4. & $1+b-a=0$ & $0$ & $V^{\prime }$ & $0$
& $0$ & $0$ & $0$ \\ \hline 5. & $a=\frac{1}{3}$ & $0$ & $0$ &
$P^{\prime }$ & $0$ & $0$ & $0$ \\ \hline 6. & $a=b=\frac{1}{3}$ &
$0$ & $0$ & $P^{\prime }$ & $0$ & $P^{\prime }$ & $0 $ \\ \hline
7. & $a+2b-1=0$ & $0$ & $0$ & $0$ & $0$ & $P^{\prime }$ & $0$ \\
\hline
\end{tabular}
\end{center}
\end{table}

\medskip
\noi We observe that among these algebras, only those for which
the condition $\left[K',W'\right]=0$ is satisfied admit the
subalgebra $\left<L'_i,K'_i,P'_i,H'\right> \oplus \left<W\right>$,
the trilinear bracket of which being trivial. Actually this subalgebra can be
considered as the cubic analogue of the static algebra of classical kinematics.
Moreover, it is straightforward to verify that this subalgebra is
always a contraction of the cubic algebra
$\left<L'_i,K'_i,P'_i,H'\right> \oplus \left<Q'\right>$ considered
in fractional supersymmetry \cite{1dfsusy}:
 \beqa \label{FSUSY} \left\{Q',Q',Q'\right\}= 3 H', \eeqa

\noi
by simply considering the transformations
\beqa \label{Kont} W=\varepsilon Q \eeqa

\noi
Although it might appear surprising that \eqref{FSUSY} does not appear as
a contraction of the cubic Poincar\'e (or de Sitter) algebra, this is a direct consequence
of the action of the Bose sector on the trilinear part and the bracket
$\left\{W,W,W\right\}$. We will encounter this obstruction to recover this special extension
also in later cases.

\section{Extensions of the Carroll algebra}
The Carroll algebra, first introduced in \cite{LL}, is obtained
from the Poincar\'e algebra through a contraction determined by
rescaling the boosts and time translations:

\beqa
\label{IW-C}
J'_i=J_i, \ \ K'_i = \epsilon K_i, \ \ P'_i = P_i, \ \ H'= \epsilon H.
\eeqa

The brackets are give on Table 1.
Although appearing naturally in the classification of kinematical
groups, as an alternative intermediate algebra in the contraction
of the Poincar\' e group onto the static group, and therefore as
another non-relativistic limit (the other being the Galilei
algebra), the Carroll algebra has played no distinguished role in
kinematics. However, recently it has been analysed whether this
algebra constitutes an object in the study of tachyon condensates
in string theory \cite{Gib}. It is in this context where a
possible cosmological interpretation of this limit of the
Poincar\'e algebra and the supersymmetric extensions recover some
interest.
First, it is rather straightforward to construct a supersymmetric
extension of the Carroll algebra, by adding to the generators
specified in \eqref{IW-C} the additional generators of the
symmetric part
\beqa
\label{supergal}
Q'_\alpha = \epsilon^a Q_\alpha, \ \ \bar Q'_\alpha = \epsilon^b \bar Q_\alpha.
\eeqa

\noi These transformations define a contraction if the constraint
$a+b-1 \ge 0$ is satisfied. For the special case $a+b-1=0$ we get
the $N=2$ supersymmetric extension of the Carroll algebra
determined by
\beqa
\label{suoer-Car}
\begin{array}{cc}
\begin{array}{cc}
\left[L'_k, Q'_\alpha\right]= -\frac{i}{2}
(\sigma_k)_\alpha{}^\beta Q'_\beta,& \left[K'_k,
Q'_\alpha\right]=0, \\
\left[L'_k, \bar Q'_\alpha\right]= -\frac{i}{2}
(\sigma_k)_\alpha{}^\beta \bar Q'_\beta,& \left[K'_k,
\bar Q'_\alpha\right]=0,
\end{array}
& \left\{Q'_\alpha, \bar Q'_\beta\right\}= 2 \delta_{\alpha \beta} H.
\end{array}
\eeqa

%\beqa
%\label{suoer-Car}
%\left\{Q'_\alpha, \bar Q'_\beta\right\} = 2 \delta_{\alpha \beta} H'.
%\eeqa

\noi
The remarkable fact of this extension is that in the Fermi sector
the symmetric product reproduces a Clifford algebra structure.
Indeed, this algebra corresponds to the $N=4$ supersymmetric
quantum mechanics extension of the Galilei algebra. However, in contrast
to the Galilei case, here the supercharges are in the spinor representation
of the rotation group.

\subsection{Carroll algebras of order 3}

Following the same ansatz as before, we construct a Carroll
algebra of order three adding new generators, in this case
\beqa \label{IW-3car} V'_i=\epsilon^a V_i, \ \ W'=\epsilon^b W.
\eeqa

\noi If we take the limit, this leads to the following constraints
upon the parameters
\beqa \label{ine-3car} a \ge 0,\ 2a+b-1 \ge 0,\ a+2b \ge 0,\ b \ge
\frac13,\ 1+b-a\ge 0,\ 1+a-b\ge 0. \eeqa

\noi
They have all to be satisfied in order to define a Lie  of order
three. It is not difficult to see that seven independent solutions
to these conditions exist, given schematically in Table 4.

\begin{table}[h]
\caption{Carroll algebras of order three}
\begin{center}
\begin{tabular}{|l||l|c|c|c|c|c|c|}
\hline &  & $\left[ K^{\prime },V^{\prime }\right] $ & $\left[
K^{\prime },W^{\prime }\right] $ & $\left\{ V^{\prime },V^{\prime
},V^{\prime
}\right\} $ & $\left\{ V^{\prime },V^{\prime },W^{\prime }\right\} $ & $%
\left\{ V^{\prime },W^{\prime },W^{\prime }\right\} $ & $\left\{
W^{\prime },W^{\prime },W^{\prime }\right\} $ \\ \hline 1. &
$a=0,b=1$ & $W^{\prime }$ & $0$ & $P^{\prime }$ & $H^{\prime }$ &
$0$ & $0$ \\ \hline 2. & $1+a-b=0$ & $W^{\prime }$ & $0$ & $0$ &
$0$ & $0$ & $0$ \\ \hline 3. & $a=\frac{4}{3},b=\frac{1}{3}$ & $0$
& $V^{\prime }$ & $0$ & $0$ & $0$ & $H^{\prime }$ \\ \hline 4. &
$1+b-a=0$ & $0$ & $V^{\prime }$ & $0$ & $0$ & $0$ & $0$ \\ \hline
5. & $2a+b-1=0$ & $0$ & $0$ & $0$ & $H^{\prime }$ & $0$ & $0$ \\
\hline 6. & $a=b=\frac{1}{3}$ & $0$ & $0$ & $0$ & $H^{\prime }$ &
$0$ & $H^{\prime } $ \\ \hline 7. & $b=\frac{1}{3}$ & $0$ & $0$ &
$0$ & $0$ & $0$ & $H^{\prime }$ \\ \hline
\end{tabular}
\end{center}
\end{table}

\medskip

\noi For all the cubic extensions  the brackets $\left\{
V^{\prime },W^{\prime },W^{\prime }\right\} $ vanishes. Among these
solutions, the last is of special interest, since it admits the
subalgebra $\left<L'_i,K'_i,P'_i,H'\right> \oplus \left<W'\right>$
and corresponds to the  fractional supersymmetry extending the
Galilei algebra
$$
\left\{W',W',W'\right\}= 3 H'.$$

\section{Extensions of the Newton algebra}
In contrast to the previous cases, the Newton-Hooke algebras are
not contractions of the Poincar\'e algebra, but are obtained
directly from the de Sitter algebras by speed-space
contractions,\footnote{These further have the Galilei algebra as
non-relativistic limit. See Table 1.} determined by the
transformations
\beqa \label{IW-N} P'_i=\epsilon P_i, \ \ K'_i=\epsilon K_i.
\eeqa

\noi The limit $\epsilon \to 0$ corresponds to the oscillating
Newton Lie algebra. The various brackets are given on Table 1. The
features of these models are therefore close to those of the de
Sitter algebras, and have been studied in connection to certain
properties of space-time curvature. Moreover, these algebras are
known to appear as subalgebras of multi-temporal conformal
algebras, and have been applied to non-relativistic branes
\cite{Her}.
The $N=2$ supersymmetric extension of the Newton algebra is
obtained by joining to the generators of \eqref{IW-N} the
additional elements
\beqa
\label{IW-sN}
Q_\alpha= \epsilon^a S_\alpha, \ \ \bar Q_\alpha= \epsilon^b \bar S_\alpha
\eeqa

\noi Performing the brackets it follows that the action of $P'$ on
$Q$ and $\bar Q$ is trivial, although the anti-commutators lead
to the constraint $a+b-1 \ge 0$. If we take the equality $a+b-1=0$, which is the
only possibility of obtaining a non-trivial symmetric bracket, we get

\beqa
\begin{array}{cc}
\begin{array}{cc}
\left[L'_k, Q'_\alpha\right]= -\frac{i}{2}
(\sigma_k)_\alpha{}^\beta Q'_\beta,& \left[K'_k,
Q'_\alpha\right]=0, \\
\left[L'_k, \bar Q'_\alpha\right]= -\frac{i}{2}
(\sigma_k)_\alpha{}^\beta \bar Q'_\beta,& \left[K'_k,
\bar Q'_\alpha\right]=0,
\end{array}
& \left\{Q'_\alpha, \bar Q'_\beta\right\}= 2
\sigma_{i}{}_{\alpha \beta} P^{\prime i},
\end{array}
\eeqa

%\beqa \left\{Q_\alpha. \bar Q_\beta\right\} = 2
%\sigma_i{}_{\alpha \beta} P^{\prime i}, \eeqa

\noi which provides the desired extension. We observe that
supersymmetric extensions of the (oscillating) Newton algebras
have already been considered in \cite{Hus}.

\subsection{Newton algebras of order 3}
To study the possible contractions of the $\mathfrak{so}(2,3)-$
algebra of order 3 onto algebras of order 3 containing the Newton
algebra in its bosonic part, we introduce the following notations
that will facilitate the analysis:

\beqa
A_{MN} = \left\{ \begin{array}{ll}A_k=A_{ij}\
 (i,j,k \text{ perm.}), & B_i= A_{0i},\\
                                  C_i=A_{i4},& R=A_{04}. \\
\end{array}
\right.
\eeqa

\noi Over this basis we have

\beqa
\label{ortho4.1}
\begin{array}{llll}
\left[L_i,A_j\right]= A_k& \left[L_i,B_j\right]= B_k&
\left[L_i,C_j\right]=C_k&
\left[L_i,R\right]=0 \\
\left[K_i, A_j\right]= B_k& \left[K_i,B_j\right]=-  A_{k}&
\left[K_i,C_j\right]= -\delta_{ij} R&
\left[K_i,R\right]=-Q_i \\
\left[P_i, A_j\right]= C_k& \left[P_i,B_j\right]=\delta_{ij} R&
\left[P_i,C_j\right]= - A_{k}&
\left[P_i,R\right]=B_i \\
\left[H,A_i\right]=0 & \left[H,B_i\right]=C_i&
\left[H,C_i\right]=-B_i& \left[H,R\right]=0
\end{array}
\eeqa

\noi
and for the trilinear brackets

\beqa
\label{orth4.2}
&&\hskip .2truecm \left\{A_i,A_j,A_k\right\}=-\delta_{ij} L_k -\delta_{ik} L_j -\delta_{jk} L_i,
\nonumber \\
&&\begin{array}{lll}
\left\{A_i,A_j,B_k\right\}= -\delta_{ij} K_k,&
\left\{A_i,A_j,C_k\right\}= -\delta_{ij} P_k,&
\left\{A_i,A_j,R\right\}= -\delta_{ij}H,
\\
\left\{A_i,B_j,B_k\right\}= -\delta_{jk} L_i,&
\left\{A_i,C_j,C_k\right\}= -\delta_{jk} L_i,&
\left\{A_i,R,R\right\}=  L_i, \\
\left\{A_i,C_j,B_k\right\}=0,& \left\{A_i,C_j,R\right\}=0,&
\left\{A_i,B_j,R\right\}=0,
\end{array} \nonumber \\
&& \hskip .2truecm  \left\{B_i,B_j,B_k\right\}=-\delta_{ij} K_k
-\delta_{ik} K_j -\delta_{jk} K_i,\
\left\{B_i,B_j,C_k\right\}=-\delta_{ij} P_k,
\\
&&\begin{array}{llll}  \left\{B_i,B_j,R\right\}=-\delta_{ij} H,&
\left\{B_i,C_j,C_k\right\}= -\delta_{jk} K_i,&
\left\{B_i,R,R\right\}=K_i,& \left\{B_i,C_j,R\right\}=0,
\end{array}\nonumber  \\
&& \hskip .2truecm \left\{C_i,C_j,C_k\right\}=-\delta_{ij} P_k-\delta_{ik}P_j -\delta_{jk} P_i,\nonumber \\
&&\begin{array}{lll} \left\{C_i,C_j,R\right\}=-\delta_{ij} H,&
\left\{C_i,R,R\right\}=P_i,& \left\{R,R,R\right\}=3 H.
\end{array} \nonumber
\eeqa

\noi
Performing the contraction defined by the scaling transformations

\beqa \label{IW-N3} A'_i=\varepsilon^a A_i, B'_i=\varepsilon^b
B_i, C'_i=\varepsilon^c C_i, R'_i=\varepsilon^d R_i, \eeqa

\noi together with \eqref{IW-N}, the limit $\varepsilon \to 0$
gives rise to a consistent contraction pattern if  the following
constraints are satisfied: $ b=c $ and the system of inequalities
\beqa &&e_1=a \ge 0, e_2=2a+b-1 \ge 0,
 e_3=2a+d \ge 0,
 e_4=a+2b \ge 0,
 e_5=a+2d \ge 0,
\nonumber \\
 &&
 e_6=b-1/3 \ge 0,
 e_7=2b+d \ge 0,
e_8=b+2d-1 \ge 0,
e_9=d,
 e_{10}=1+a-b \ge 0,
\nonumber \\
&& e_{11}=1+b-a \ge 0,
 e_{12}=1+b-d \ge 0,
 e_{13}=1+d-b \ge 0.
\nonumber
\eeqa

\noi To simplify the presentation we write the non-zero brackets
in matrix form as follows
\[
\mathbf{M=}\left(
\begin{array}
[c]{cccc}%
\left[  L',A'\right]   & \left[  L',B'\right]   & \left[
L',C'\right] & \left[
L',R'\right]  \\
\left[  K',A'\right]   & \left[  K',B'\right]   & \left[
K',C'\right] & \left[
K',R'\right]  \\
\left[  P',A'\right]   & \left[  P',B'\right]   & \left[
P',C'\right] & \left[
P',R'\right]  \\
\left[  H',A'\right]   & \left[  H',B'\right]   & \left[
H',C'\right] & \left[ H',R'\right]
\end{array}
\right)  =\left(
\begin{array}
[c]{cccc}%
A' & B' & C' & 0\\
\varepsilon^{1+a-b}B' & \varepsilon^{1+b-a}A' & \varepsilon^{1+c-d}R' &
\varepsilon^{1+d-c}C'\\
\varepsilon^{1+a-c}C' & \varepsilon^{1+b-d}R' &
\varepsilon^{1+c-a}A' &
\varepsilon^{1+d-b}B'\\
0 & \varepsilon^{b-c}C' & \varepsilon^{c-b}B' & 0
\end{array}
\right)
\]%

\noi
for the bilinear part
\begin{align*}
\mathbf{N} &  =\left(
\begin{array}
[c]{cccc}%
\left\{  A',A',A'\right\}   & \left\{  A',A',B'\right\} & \left\{
A',A',C'\right\}
& \left\{  A',A',R'\right\}  \\
\left\{  A',B',B'\right\}   & \left\{  A',C',C'\right\} & \left\{
A',R',R'\right\}
& \left\{  B',B',B'\right\}  \\
\left\{  B',B',C'\right\}   & \left\{  B',B',R'\right\} & \left\{
B',C',C'\right\}
& \left\{  B',R',R'\right\}  \\
\left\{  C',C',C'\right\}   & \left\{  C',C',R'\right\} & \left\{
C',R',R'\right\} & \left\{  R',R',R'\right\}
\end{array}
\right)  \\
&  =\left(
\begin{array}
[c]{cccc}%
\varepsilon^{3a}L' & \varepsilon^{2a+b-1}K' & \varepsilon^{2a+c-1}P' &
\varepsilon^{2a+d}H'\\
\varepsilon^{a+2b}L' & \varepsilon^{a+2c}L' & \varepsilon^{a+2d}L' &
\varepsilon^{3b-1}K'\\
\varepsilon^{2b+c-1}P' & \varepsilon^{2b+d}H' & \varepsilon^{b+2c-1}K' &
\varepsilon^{b+2d-1}K'\\
\varepsilon^{3c-1}P' & \varepsilon^{2c+d}H' & \varepsilon^{c+2d-1}P' &
\varepsilon^{3d}H'
\end{array}
\right)
\end{align*}
for the trilinear part. Solving the system, we get 27 possible
algebras. All the solutions have the following brackets in common:
$[L',A']=A', \ [L',B']=B', \ [L',C']=C', \ [H',B']=C', \
[H',C']=B'$. However, among these 27 algebras, only 19 are of real
interest for providing non-trivial trilinear brackets. We
therefore only list the solutions with non-vanishing ternary
brackets. (The non-listed solutions correspond to the cases where
the identity $e_i$ are indeed an equality for (1) $i=10$, (2)
$i=11$,  (3) $i=12$, (4) $i=13$, (5) $i=10,12$, (6) $i=10,13$,(7)
$i=11,12$,(8) $i=11,13$ for which only some bilinear brackets are
non-vanishing.) In all the solutions listed in Table 5, since
$[H',C']=B'$, we do not have a solution analogous to fractional
supersymmetric extension of the Galilei algebra \eqref{FSUSY}
obtained from the Carroll algebra. This follows from the structure
of the Newton algebras, where the time translation interchanges
the space translations and the boosts, and is in perfect agreement
with the fact that contracting the Poincar\'e algebra of order
three does not lead to this exceptional cubic Galilean extension.

\begin{sidewaystable}[p]
\caption{Possible 3-algebras with Bose sector of Newton type and
non-trivial ternary brackets.}
\begin{tabular}{|c|c|c|c|c|c|c|c|c|c|c|c|c|c|c|c|c|c|c|c|}
\hline & $S1$ & $S2$ & $S3$ & $S4$ & $S5$ & $S6$ & $S7$ & $S8$ &
$S9$ & $S10$ &  $S11$ & $S12 $ & $S13$ &  $S14$ & $S15$ &
$S16$ & $S17$ & $S18$ & $S19$ \\
$e_{i}=0$:\footnotemark[1] & $_{2}$ & $_{6}$ & $_{8}$ &
$_{1..3,5}$ & $_{1,2,10}$ & $_{1,2,10,12}$ & $_{2,6,8}$ & $_{2,6}$
& $_{2,6,12}$ & $_{6,8}$ & $_{6,12}$ & $_{6,8,11}$ & $_{6,11,12}$
& $_{6,11}$ & $_{2,8}$ &
$_{2,12}$ & $_{8,9,13} $ & $_{8,9,11,13}$ & $_{8,11}$\\
 &  &  & & $_{8..10,13}$  &  &
 &  &  &  &  &  &  &  &  &  &  &  & & \\\hline
$\lbrack K^{\prime },A^{\prime }]$ & $0$ & $0$ & $0$ & $B^{\prime }$ & $%
B^{\prime }$ & $B^{\prime }$ & $0$ & $0$ & $0$ & $0$ & $0$ & $0$ &
$0$ & $0$ & $0$ & $0$ & $0$ & $0$ & $0$ \\ \hline
$\lbrack
K^{\prime },B^{\prime }]$ & $0$ & $0$ & $0$ & $0$ & $0$ & $0$ &
$0$ & $0$ & $0$ & $0$ & $0$ & $A^{\prime }$ & $A^{\prime }$ &
$A^{\prime }$ & $0$ & $0$ & $0$ & $A^{\prime }$ & $A^{\prime }$ \\
\hline
$\lbrack K^{\prime },C^{\prime }]$ & $0$ & $0$ & $0$ & $0$ & $0$ & $%
R^{\prime }$ & $0$ & $0$ & $R^{\prime }$ & $0$ & $R^{\prime }$ & $0$ & $%
R^{\prime }$ & $0$ & $0$ & $R^{\prime }$ & $0$ & $0$ & $0$ \\
\hline
$\lbrack K^{\prime },R^{\prime }]$ & $0$ & $0$ & $0$ &
$C^{\prime }$ & $0$ & $0$ & $0$ & $0$ & $0$ & $0$ & $0$ & $0$ &
$0$ & $0$ & $0$ & $0$ & $C^{\prime }$ & $C^{\prime }$ & $0$ \\
\hline
$\lbrack P^{\prime },A^{\prime }]$ & $0$ & $0$ & $0$ & $C^{\prime }$ & $%
C^{\prime }$ & $C^{\prime }$ & $0$ & $0$ & $0$ & $0$ & $0$ & $0$ &
$0$ & $0$ & $0$ & $0$ & $0$ & $0$ & $0$ \\ \hline
$\lbrack P^{\prime },B^{\prime }]$ & $0$ & $0$ & $0$ & $0$ & $0$ & $%
R^{\prime }$ & $0$ & $0$ & $R^{\prime }$ & $0$ & $R^{\prime }$ & $0$ & $%
R^{\prime }$ & $0$ & $0$ & $R^{\prime }$ & $0$ & $0$ & $0$ \\
\hline $\lbrack P^{\prime },C^{\prime }]$ & $0$ & $0$ & $0$ & $0$
& $0$ & $0$ & $0$ & $0$ & $0$ & $0$ & $0$ & $A^{\prime }$ &
$A^{\prime }$ & $A^{\prime }$ & $0$ & $0$ & $0$ & $A^{\prime }$ &
$A^{\prime }$ \\ \hline $\lbrack P^{\prime },R^{\prime }]$ & $0$ &
$0$ & $0$ & $B^{\prime }$ & $0$ & $0$ & $0$ & $0$ & $0$ & $0$ &
$0$ & $0$ & $0$ & $0$ & $0$ & $0$ & $B^{\prime }$ & $B^{\prime }$
& $0$ \\
\hline $\{A^{\prime },A^{\prime },A^{\prime }\}$ & $0$ &
$0$ & $0$ & $L^{\prime }$
& $L^{\prime }$ & $L^{\prime }$ & $0$ & $0$ & $0$ & $0$ & $0$ & $0$ & $0$ & $%
0$ & $0$ & $0$ & $0$ & $0$ & $0$ \\
\hline
$\{A^{\prime },A^{\prime },B^{\prime }\}$ & $K^{\prime }$ & $0$ & $0$ & $%
K^{\prime }$ & $K^{\prime }$ & $K^{\prime }$ & $K^{\prime }$ &
$K^{\prime }$
& $K^{\prime }$ & $0$ & $0$ & $0$ & $0$ & $0$ & $K^{\prime }$ & $%
K^{\prime }$ & $0$ & $0$ & $0$ \\
\hline
$\{A^{\prime },A^{\prime },C^{\prime }\}$ & $P^{\prime }$ & $0$ & $0$ & $%
P^{\prime }$ & $P^{\prime }$ & $P^{\prime }$ & $P^{\prime }$ &
$P^{\prime }$
& $P^{\prime }$ & $0$ & $0$ & $0$ & $0$ & $0$ & $P^{\prime }$
 & $%
P^{\prime }$ & $0$ & $0$ & $0$ \\ \hline $\{A^{\prime },A^{\prime
},R^{\prime }\}$ & $0$ & $0$ & $0$ & $H^{\prime }$ & $0$ & $0$ &
$0$ & $0$ & $0$ & $0$ & $0$ & $0$ & $0$ & $0$ & $0$ & $0$ & $0$ &
$0$ & $0$ \\ \hline $\{A^{\prime },B^{\prime },B^{\prime }\}$ &
$0$ & $0$ & $0$ & $0$ & $0$ & $0$ & $0$ & $0$ & $0$ & $0$ & $0$ &
$0$ & $0$ & $0$ & $0$ & $0$ & $0$ & $0$ & $0$
\\ \hline
$\{A^{\prime },C^{\prime },C^{\prime }\}$ & $0$ & $0$ & $0$ & $0$
& $0$ & $0$ & $0$ & $0$ & $0$ & $0$ & $0$ & $0$ & $0$ & $0$ & $0$
& $0$ & $0$ & $0$ & $0$
\\ \hline
$\{A^{\prime },R^{\prime },R^{\prime }\}$ & $0$ & $0$ & $0$ &
$L^{\prime }$ & $0$ & $0$ & $0$ & $0$ & $0$ & $0$ & $0$ & $0$ &
$0$ & $0$ & $0$ & $0$ & $0$ & $0$ & $0$ \\
\hline $\{B^{\prime
},B^{\prime },B^{\prime }\}$ & $0$ & $K^{\prime }$ & $0$ & $0$ &
$0$ & $0$ & $K^{\prime }$ & $K^{\prime }$ & $K^{\prime }$ &
$K^{\prime }$ & $K^{\prime }$ & $K^{\prime }$ & $K'$ & $K^{\prime
}$ & $0$ & $0$ & $0$ & $0$ & $0$ \\
\hline $\{B^{\prime
},B^{\prime },C^{\prime }\}$ & $0$ & $P^{\prime }$ & $0$ & $0$ &
$0$ & $0$ & $P^{\prime }$ & $P^{\prime }$ & $P^{\prime }$ &
$P^{\prime }$ & $P^{\prime }$ & $P^{\prime }$ & $P'$ & $P^{\prime
}$ & $0$ & $0$ & $0$ & $0$ & $0$ \\
 \hline $\{B^{\prime
},B^{\prime },R^{\prime }\}$ & $0$ & $0$ & $0$ & $0$ & $0$ & $0$
& $0$ & $0$ & $0$ & $0$ & $0$ & $0$ & $0$ & $0$ & $0$ & $0$ & $0$ & $%
K^{\prime }$ & $0$ \\
\hline $\{B^{\prime },C^{\prime },C^{\prime
}\}$ & $0$ & $K^{\prime }$ & $0$ & $0$ & $0$ & $0$ & $K^{\prime }$
& $K^{\prime }$ & $K^{\prime }$ & $K^{\prime }$ & $K^{\prime }$ &
$K^{\prime }$ & $K'$ & $K^{\prime }$ & $0$ & $0$ & $0$ & $0$ & $0$
\\ \hline
$\{B^{\prime },R^{\prime },R^{\prime }\}$ & $0$ & $0$ & $K^{\prime }$ & $%
K^{\prime }$ & $0$ & $0$ & $K^{\prime }$ & $0$ & $0$ & $K^{\prime
}$ & $0$ &
$K^{\prime }$ & $0 $ & $0$ & $K^{\prime }$ & $0$ & $K^{\prime }$ & $%
0$ & $K^{\prime }$ \\
\hline
$\{C^{\prime },C^{\prime },C^{\prime
}\}$ & $0$ & $P^{\prime }$ & $0$ & $0$ & $0$ & $0$ & $P^{\prime }$&
 $P^{\prime }$ & $P^{\prime }$ & $P^{\prime }$ & $P^{\prime }$ &
$P^{\prime }$ & $P' $ & $P^{\prime }$ & $0$ & $0$ & $0$ & $0$ & $0$
\\
\hline $\{C^{\prime },C^{\prime },R^{\prime }\}$ & $0$ & $0$ &
$0$ & $0$ & $0$ & $0$ & $H^{\prime }$ & $0$ & $0$ & $0$ & $0$ &
$0$ & $0$ & $0$ & $0$ & $0$ & $0$ & $0$ & $0$ \\ \hline
$\{C^{\prime },R^{\prime },R^{\prime }\}$ & $0$ & $0$ & $P^{\prime }$ & $%
P^{\prime }$ & $0$ & $0$ & $0$ & $0$ & $0$ & $P'$ & $0$ & $%
P^{\prime }$ & $0 $ & $0$ & $P^{\prime }$ & $0$ & $P^{\prime }$ & $%
P^{\prime }$ & $P^{\prime }$ \\ \hline $\{R^{\prime },R^{\prime
},R^{\prime }\}$ & $0$ & $0$ & $0$ & $H^{\prime }$
& $0$ & $0$ & $0$ & $0$ & $0$ & $0$ & $0$ & $0$ & $0$ & $0$ & $0$ & $0$ & $%
H^{\prime }$ & $H^{\prime }$ & $0$ \\ \hline
\end{tabular}
\footnotetext[1]{For any solution $S$ the indices make reference
to the inequalities of the system that are identities. For the
non-appearing indices, the corresponding equations are strict
inequalities}
\end{sidewaystable}

\subsection{Contractions of cubic Newton extensions}

Since the Newton algebra contracts onto the Galilei algebra by
means of the
transformations%
\beqa
 P^{\prime}=\varepsilon P,\;H^{\prime}=\varepsilon H, \label{NG}
\eeqa it is expected that the transitivity of contractions implies
that the cubic extensions of Newton contract onto some cubic
extensions of the Galilei algebra (of dimension 20).  It is
straightforward to verify that choosing \beqa
A^{\prime}=\varepsilon A,\; B^{\prime}=\varepsilon B,\;
C^{\prime}=\varepsilon^{\frac{1}{3}}C,\;R^{\prime}=\varepsilon^{\frac{1}{3}}R,
\eeqa the trilinear brackets of the four elements
$\left\{C_{i},R\right\}$ is preserved. Further these
transformations imply that, after the contraction, $P'$ and $H'$
commute with $A',B',C',R'$. As a consequence, all the contractions
admit a subalgebra spanned by the generators
$<L_{i},K_{i},P_{i},H>\oplus<C\,_{i},R>$, which therefore must
correspond to some cubic extensions $E_{i}$ ($i=1..7)$ of the
Galilei algebra obtained in Table 3. Denoting by $S_{j}$ the
algebras of Table 5, the preceding transformations give rise to
the following contraction pattern:

\begin{enumerate}
\item $E_{1}$ arises from the contraction of $S_{9}$, $S_{11}$ and
$S_{13}$.

\item $E_{2}$ arises from the contraction of $S_{16}.$

\item $E_3$ arises from the contraction of $S_{4}$, $S_{17}$ and
$S_{18}$.

\item $E_{5}$ arises from the contraction of $S_{2}$, $S_{8}$ and
$S_{14}$.

\item $E_{6}$ arises from the contraction of $S_{7}$,  $S_{10}$
and $S_{12}$.

\item $E_{7}$ arises from the contraction of $S_{3}$ and $S_{15}$
and $S_{19}$.
\end{enumerate}

\bigskip
\noi Further, $S_{1}$  $S_{5}$ and $S_{6}$ contract onto a trivial
cubic extension of the Galilei algebra\footnote{That is, having
zero bilinear and trilinear brackets.}. As for the algebra
$E_{4}$, it follows from the contraction of the algebras
$S_{4},S_{17}$ and $S_{18}$ defined by the transformations: \beqa
A^{\prime}=\varepsilon A, B^{\prime}=\varepsilon B,
C^{\prime}=\varepsilon C, R^{\prime}=\varepsilon R \eeqa in
addition to those of (\ref{NG}). This shows that any of the
Galilean algebras of order three can be obtained through two
different directions, either starting from the extension of the
Poincar\'e algebra (\ref{3poin}), or considering the Newton
algebras of order three. The remarkable fact is that in the latter
case, the Galilean extensions are obtained as subalgebras of
contractions, in contrast to the derivation developed in section
4. This points out the consistency of the contraction method, and
justifies that no information was lost in considering mainly the
algebra (\ref{3poin}) and its contractions.

\section{Conclusions}

Following the procedure undertaken by Bacry and L\'evy-Leblond to
classify the possible kinematics, we have extended the method,
based on contractions of Lie algebras, to obtain a similar
classification of ``kinematical" superalgebras and Lie algebras of
order three, basing on natural generalisations of the already
known extensions that have been proven to be of physical interest.
In most cases this is possible using the supersymmetric extension
of the anti-de Sitter algebra. This contraction pattern gives rise
to some of the Lie algebras of order three which have been used in
physics with interesting applications (see e.g. \cite{csusy}).
However, some new cubic algebras emerge naturally is this
classification scheme. Whether or not they could have some
physical interesting interpretation is still an open question. In
this paper, various extensions arising directly from the cubic de
Sitter algebras have been left out, which provide other
alternative models. However, due to their non-apparent relation
with the established field theoretic realizations, they seem not
to be very relevant from the physical point of view. The
impossibility of deriving the Galilean extension considered in
FSUSY and the supersymmetric Quantum Mechanics shows that some
possibilities are lost when generalising the contraction procedure
from Lie algebras to Lie algebras of order three. This exceptions
are however expectable from the theory, and do not constitute a
limitation of the method.

\smallskip

It should also be remarked that the classical kinematics are based
on a heavy assumption, namely, that parity and time reversal
(short PT) are automorphisms of the kinematical group. However, as
follows from the theory of weak interactions, this hypothesis is
physically objectionable. Moreover, it is known that a permanent
neutron electric dipole moment requires parity and time-reversal
violation \cite{Her1}, which invalidates partially the physical
assumptions made in \cite{bl}. A classification of kinematics
based only on spatial isotropy exists \cite{Nu}, and the resulting
algebras are also related by contractions. However, these
additional algebras have no relation with the de Sitter algebras,
that is, no contraction of the latter leads to these models
without (PT). Moreover, they may also depend on parameters, which
makes a physical interpretation of the generators quite difficult.
Due to the non-existence of some relation between the
non-classical kinematical algebras and semisimple Lie algebras,
the possibility of obtaining supersymmetric and cubic extensions
is probably very reduced, if feasible at all.

\subsection*{Acknowledgement}
One of the authors (RCS) acknowledges partial financial support
from the research project MTM2006-09152 of the Ministerio de
Educaci\'on y Ciencia. The authors are indebted to J. A. de
Azc\'arraga for useful comments and references \cite{Zu}.

\appendix

\section{Lie algebra of order $F$$-$definition}

We summarize here for completeness the definition of a (real)
elementary Lie algebra of order three, since in this paper we are
dealing with this type of algebras. More details can be found in
\cite{flie}. The real vector space $\g= \g_0 \oplus \g_1$ is
called an elementary Lie algebra of order three if

\begin{enumerate}
\item $\mathfrak{g}_0$ is a (real) Lie algebra. \item
$\mathfrak{g}_1$ is a (real) representation of
$\mathfrak{g}_0$. If $X \in \g_0, Y \in \g_1$, then $[X,Y]$
denotes the action of $X \in \g_0$ over $Y \in \g_1$.
 \item  There exists  a
 trilinear,  $\mathfrak{g}_0$-equivariant  map
 $$\{ \ ,\ , \  \} : {\cal S}^3\left(\mathfrak{g}_1\right)
\rightarrow \mathfrak{g}_0,$$
 where  ${ \cal S}^3(\mathfrak{g}_1)$ denotes
the three$-$fold symmetric product of $\mathfrak{g}_1$. \item  For
$Y_1,\cdots,Y_{4} \in \mathfrak{g}_1$ the following ``Jacobi
identity'' holds:
\beqa
\label{eq:J}
[Y_1,\{Y_2,Y_3,Y_4\}]+[Y_2,\{Y_3,Y_4,Y_1\}]+[Y_3,\{Y_4,Y_1,Y_2\}]+
[Y_4,\{Y_1,Y_2,Y_3\}]=0.
\eeqa
\end{enumerate}

\noi It turns out that
Lie algebras of order three appear as some kind of generalization
of Lie algebras and superalgebras. In \cite{flie} the definition
of  (complex) Lie algebras of order $F$ (elementary or not) has
been given. Indeed, in the complex case the situation is more
rich, since there is, in addition to the real case, an
automorphism which induces a grading of the (complex) vector
space.

\end{document}